\documentclass[onecolumn, 12pt]{IEEEtran}
\usepackage{setspace}
\ifCLASSINFOpdf
   \usepackage[pdftex]{graphicx}
   \graphicspath{{../pdf/}{../jpeg/}}
   \DeclareGraphicsExtensions{.pdf,.jpeg,.png}
\else
   \usepackage[dvips]{graphicx}
   \graphicspath{{../eps/}}
   \DeclareGraphicsExtensions{.eps}
\fi
\usepackage{epstopdf}
\usepackage{amsmath}
\usepackage{bm}

\usepackage{subfigure}
\usepackage{caption}
\usepackage{algorithm}
\usepackage{algorithmic}

\usepackage{amssymb}
\usepackage{color}

\begin{document}
\captionsetup[figure]{labelfont={default},labelformat={default},labelsep=period,name={Fig.}}
\title{Cooperative Reflection Design with Timing Offsets in Distributed Multi-RIS Communications}
%
%
%
\author{Yaqiong Zhao, \emph{Student Member, IEEE,}
       Wei Xu, \emph{Senior Member, IEEE,}
       Huan Sun, \emph{Member, IEEE,}\\
       Derrick Wing Kwan Ng, \emph{Fellow, IEEE,}
      and Xiaohu You, \emph{Fellow, IEEE}
 \thanks{Y. Zhao, W. Xu, and X. You are with the National Mobile Communications Research Lab., Southeast University, Nanjing 210096, China, and also with Purple Mountain Laboratories, Nanjing 211111, China (\{zhaoyaqiong, wxu, xhyu\} @seu.edu.cn).}
\thanks{H. Sun is with the Wireless Technology Laboratory, Huawei Technologies Co. Ltd, Shanghai 201206, China (sunhuan11@huawei.com).}
\thanks{D. W. K. Ng is with the School of Electrical Engineering and Telecommunications, the University of New South Wales, NSW 2052, Australia (w.k.ng@unsw.edu.au).}
}

\maketitle
\begin{abstract}
 This letter investigates a wireless communication system deploying distributed reconfigurable intelligent surfaces (RISs). Existing works have assumed that perfect  timing synchronization is available among all the cooperative RISs. For practical considerations, we first study cooperative reflection design for multi-RIS-aided communication systems taking into account timing synchronization errors. We aim to minimize the mean-squared error of the recovered data in the presence of timing offsets subject to the unit-modulus constraints imposed on the phase shifts of the RISs. In order to handle this sophisticated nonconvex problem, we develop a computationally-efficient algorithm based on the majorization-minimization framework where the RIS phase shift matrices and the timing offset equalizer are respectively developed in a semi-closed form and a closed form. Numerical examples validate the improved performance of our proposed design compared with various schemes.

\end{abstract}


%
\IEEEpeerreviewmaketitle
\section{Introduction}
\IEEEPARstart{R}{econfigurable} intelligent surface (RIS) has been regarded as an emerging technology to significantly improve both the link reliability and the service coverage [1]. It consists of a large number of low-cost passive elements, each of which is reconfigurable by inducing a certain phase shift to the reflected impinging signals. Hence, by carefully adapting the phase shifts to the instantaneous channel conditions, RIS is able to establish favorable wireless channels and offer a great potential in performance improvement [2].

To fully exploit this potential of RIS techniques, considerable efforts have been devoted in point-to-point communications [3], [4] and multi-user systems [5], [6]. For instance, a cascaded channel estimation method was proposed for a point-to-point RIS-aided communication system in [3]. Also, for a RIS-aided orthogonal frequency division multiplexing system, the authors maximized the system sum-rate by jointly optimizing the transmit beamforming and the RIS phase shifts [4]. As for multiuser communications, a channel estimation protocol was proposed in [5] for the RIS-aided downlink transmission. Besides, in [6], the power allocation and phase shifts were jointly optimized for RIS systems to maximize the total sum-rate of multiple users.

To pave the way for practical implementation of RIS-aided systems, these works were recently extended to systems deploying multiple RISs. For example, the sum-rate maximization problem for a multi-RIS-aided system was respectively studied for point-to-point [7] and multi-user setups [8]. Particularly, it was pointed out in [7] that a distributed system with multiple RISs outperforms the counterpart deployment of a single centralized RIS with the same number of reflecting elements, making multiple RISs more promising. On the other hand, different from the above works assuming that multiple RISs are always switched on, RIS on-off statuses were further optimized in [9] to maximize the system energy efficiency. In fact, existing works about multi-RIS over optimistically assumed perfect timing synchronization among RISs, based on which the parameters are optimized to improve the performance. Nevertheless, it is challenging to achieve perfect timing synchronization for passive RISs even with a reasonable cell size, since the RISs cannot estimate the individual timing offset of each hop of the links. In other words, the timing offsets of the multi-RIS-aidded system cannot be compensated separately by each RIS circuit as in traditional multi-hop systems. More importantly, applying the existing results based on perfect timing synchronization to practical RIS systems would jeopardize the system performance. Thus, there is an emerging need for the design of efficient RIS which takes into account the performance degradations due to asynchronization.

To address this issue, we investigate cooperative reflection design for a multi-RIS-aided system in the presence of timing offsets. The design goal is to minimize the mean-squared error (MSE) of the recovered data by jointly optimizing the RIS phase shift matrices and the timing offset equalizer considering timing synchronization errors. The formulated problem is nonconvex with the unit-modulus constraint imposed on the RIS phase shifts. To acquire a tractable solution, we devise an efficient suboptimal algorithm based on the majorization-minimization (MM) technique. The original problem is approximated in each iteration by a convex problem which admits a closed-form solution. The proposed algorithm is guaranteed to converge to a local optimum of the original design problem. Simulation results unveil
the performance superiority of the proposed method compared to various baselines.

The rest of this paper is organized as follows. System model is introduced in Section \uppercase\expandafter{\romannumeral2}. In Section \uppercase\expandafter{\romannumeral3}, we devise the RIS phase shift matrices jointly with the timing offset equalizer. Simulation results and conclusions are given in Section \uppercase\expandafter{\romannumeral4} and Section \uppercase\expandafter{\romannumeral5}, respectively.

Notation: diag$\left(  \cdot  \right)$ returns a diagonal matrix with the input as its elements and blk$\left[{\bf A},\cdots,{\bf B}\right]$ denotes the block-diagonal matrix with ${\bf A},\cdots,{\bf B}$ on its diagonal. Re$\{\cdot\}$ takes the real part of a complex quantity and vec$(\cdot)$ stands for the vectorization operation. The operation tr$(\cdot)$ takes the trace of the matrix and  $\mathbb{E}\{\cdot\}$ takes the expectation of a random variable. ${\bf I}_K$ denotes a $K\times K$ identity matrix. Notation $\|\cdot\|_1$ and $\|\cdot\|_{F}$, respectively, stand for the $L_1$ norm and the Frobenius norm of the matrix, while $\lambda_{\text{max}}(\cdot)$ being the maximum eigenvalue of the input matrix. ${\cal {CN}}(0,\sigma^2)$ represents the zero-mean complex Gaussian distribution with zero mean and the variance $\sigma^2$. Finally, $\otimes$ stands for the Kronecker product.
\section{System Model}
We consider a distributed multiple RIS downlink system as illustrated in Fig. 1, where $K$ distributed RISs are deployed to assist the data transmission from a single-antenna source, ${\mathbb S}$, to a single-antenna destination, ${\mathbb D}$, under the control of the RIS controller. Each of the RISs is equipped with $N$ passive reflecting elements. The baseband channel from ${\mathbb D}$ to the $k$th RIS, $\forall k\in\{1,\cdots,K\},~{\mathbb I}_k$, and the channel from ${\mathbb S}$ to ${\mathbb I}_k$, are denoted by ${\bf h}_k\in {\mathbb C}^{N\times 1}$ and ${\bf f}_k\in {\mathbb C}^{N\times 1}$, respectively. ${\bf W}_k= \text {diag}({\bm \theta _k}) \in {\mathbb C}^{N\times N}$ stands for the diagonal phase shift matrix for ${\mathbb I}_k$ with ${\bm \theta _k}$ being the vector of phase shifts introduced by ${\mathbb I}_k$. Direct links between ${\mathbb S}$ and ${\mathbb D}$ are assumed weakly enough to be neglected due to severe blocked propagation conditions, as adopted in the literature, e.g., [1], [3], [8].

In practice, due to hardware inconsistencies and distributed locations of RISs, the signals from ${\mathbb S}$ arriving at ${\mathbb D}$ experience different propagation delays through the distributed RISs. In other words, the received signals from different RISs are generally asynchronized. We denote the normalized timing offset of ${\mathbb I}_k$ over the symbol duration $T$ by $\epsilon_k$. Note that $\epsilon_k$ is the cascaded timing offset which consists of the timing offset of the ${\mathbb S}$-${\mathbb I}_k$ link and that of the ${\mathbb I}_k$-${\mathbb D}$ link. Mathematically, the received signal (within $0\leq t\leq L_0T$) at ${\mathbb D}$ is expressed as
\begin{equation}\label{eq:timedomin}
y(t)=\!\sum_{k=1}^{K}{\bf h}_k^H{\bf W}_k{\bf f}_k\sum_{i=-L_g}^{L_0+L_g-1}\!s(i) g\left(t-i T-\epsilon_{k} T\right)+v(t),
\end{equation}
where $g(t)$ is the transmitter pulse shaping filter and $s(i)$ is the complex-valued symbol. The last term $v(t)$ is the zero-mean, circularly symmetric complex additive white Gaussian noise with variance $\sigma^2$. Here, symbol $L_0$ represents the observation interval while $L_g$ is the transmitter pulse shaping filter lag.

Upon reception, the signal is oversampled by a factor $Q=T/T_s$, where $T_s$ is the sampling period. Note that the length of sequence ${\bf s}$ transmitted from ${\mathbb S}$ is defined as $L=2L_g+L_0$, i.e., ${\bf s}\triangleq[s(0),\cdots,s(L-1)]^T$, according to (\ref{eq:timedomin}). By stacking the $L_0Q$ received samples, i.e., ${\bf y}\triangleq[y(0),y(T_s),\cdots,y(L_0Q-1)T_s]^T$, the received vector in (\ref{eq:timedomin}) becomes
\begin{equation}\label{eq:vector}
{\bf y}=\sum_{k=1}^{K} \left({\bf h}_k^{\text H}{\bf W}_k{\bf f}_{k}\right){\bf A}_{\epsilon_{k}} {\bf s}+{\bf v},
\end{equation}
where ${\bf v}\triangleq[v(0),\!\cdots\!,v(L_0Q-1)T_s]^T$, \!$\mathbf{A}_{\epsilon_{k}}\!\triangleq\!\left[\mathbf{a}_{-L_g}\left(\epsilon_{k}\right), \cdots, \mathbf{a}_{L_0+L_g-1}\left(\epsilon_{k}\right)\right]$, and $\mathbf{a}_{i}\left(\epsilon_{k}\right)\triangleq\left[g\left(-i T-\epsilon_{k} T\right),\right.$
$\left.\cdots,g\left(-i T+\left(L_0Q-1\right) T_{s}-\epsilon_{k}T\right)\right]^{T}.$
 Now by defining $\mathbf{A}_{\epsilon}\triangleq\left[\mathbf{A}_{\epsilon_{1}}, \cdots, \mathbf{A}_{\epsilon_{K}}\right]$, $\mathbf{H}\triangleq\text{blk}\left[\mathbf h_{1}^{H}, \cdots, \mathbf h_{K}^{H}\right] \otimes{\mathbf I_L}$, $\mathbf{F} \triangleq\left[\mathbf f_{1}^{T}, \cdots, \mathbf f_{K}^{T}\right]^{T} \otimes{\mathbf I_L}$, and $\mathbf{W} \triangleq \text{blk}\left[\mathbf W_{1}, \cdots, \mathbf W_{K}\right] \otimes{\mathbf I_L}$, the equation in (\ref{eq:vector}) can be written in a concise form as
\begin{equation}\label{eq:matrix}
\mathbf{y}=\mathbf{A}_{\epsilon}\mathbf{H}\mathbf{W}\mathbf{F}\mathbf{s}+\mathbf{v}.
\end{equation}

In distributed RIS systems, the reflected signals are no longer perfectly aligned in time with each other at the destination, giving rise to severe inter-symbol interference (ISI). As a result, the diversity gain achieved by optimized RIS reflection and beamforming is reduced due to the timing misalignment. In the following, we design the RIS phase shift matrices jointly with the timing offset equalizer $\bf G$ at ${\mathbb D}$ by taking this timing asynchronization into account for enhancing the performance.
\begin{figure}[!t]
\setlength{\abovecaptionskip}{0pt}
\setlength{\belowcaptionskip}{0pt}
\centering 
\includegraphics[width=3.0in]{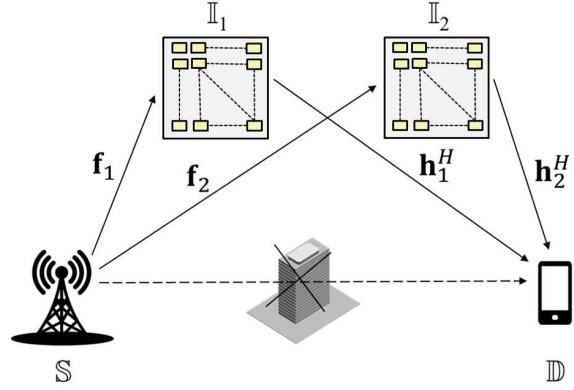} 
\caption*{\small{Fig. 1 System model for a multi-RIS-aided system with $K=2$ RISs.}}
\label{fig:0}       
\end{figure}
\section{Cooperative Reflection and Synchronization Design}
\subsection{Problem Reformulation}
To improve the tractability of (\ref{eq:matrix}), we start with an equivalent transformation of (\ref{eq:matrix}) by applying the change of variables $\mathbf{H}\mathbf{W}\mathbf{F}=\bf{H}_{\text {eq}}{\bf{\Theta }}$, where ${\bf H}_{\text{eq}}\triangleq {\text{blk}}\left[ {{\bf{h}}_1^H{\text{diag}}\left( {{{\bf{f}}_1}} \right), \cdots ,{\bf h}_K^H{\text{diag}}\left( {{{\bf{f}}_K}} \right)} \right] \otimes {{\bf{I}}_L}$ and ${\bm{\Theta }} \triangleq\bm\theta\otimes {{\bf{I}}_L}$, where $\bm\theta\triangleq{\left[ {\bm \theta _1^T, \cdots ,\bm \theta _K^T} \right]^T}$. Then (\ref{eq:matrix}) can be rewritten as
\begin{equation}\label{eq:convert}
{\bf{y}}=\bf{A}_{\epsilon}\bf{H}_{\text {eq}}{\bf{\Theta }}\bf{s}+\bf{v}.
\end{equation}
Note that $\bf{H}_{\text {eq}}$ and $\epsilon_k$ are assumed to be obtained through joint channel and timing estimation by exploiting the ``ON/OFF" scheme like in [5] and the least squares estimator [10]. Given the information of $\bf{H}_{\text {eq}}$, $\epsilon_k$, and (\ref{eq:convert}), we propose to jointly optimize $\bm\Theta$ and $\bf G$ taking into account the asynchronization to minimize the recovered data MSE under the constant-modulus constraint which yields:
\begin{subequations}\label{eq:problem}
\begin{align}
&{\mathop{\text{minimize}}\limits_{_{{\bm\Theta} ,{\bf G}}}}\quad\text{MSE}({\bm\Theta},{\bf G})\triangleq\mathbb{E}\left\{ {\left\|{\bf {Gy}} - {\bf T}({\bm{\eta }}){\bf s} \right\|^2 }\right\}\label{eq:problema}\\
&\quad\quad\quad\text{s.t.}\quad{\bm\Theta}={\bm\theta}\otimes{\bf I}_L,\label{eq:problemb}\\
&\quad\quad\quad\quad\quad\left|\theta_i\right|=1,\quad\forall i=1,\cdots,NK,\label{eq:problemc}
\end{align}
\end{subequations}
where $\theta_i$ denotes the $i$th element of $\bm\theta$, and ${\bf T}(\bm{\eta })$ is an $L_0\times L$
circulant matrix which can be perceived as a windowing operation selecting the length-$L_0$ block of data for detection [10]. The first row of the circulant ${\bf T}(\bm{\eta })$ gives as
\begin{equation}\label{eq:eta}
{\bm\eta} \triangleq\left[{\bf R}_{\text {g}}\left(-L_g\right), \cdots, {\bf R}_{\text {g}}(0), \cdots, {\bf R}_{\text {g}}\left(L_{g}\right) ~ \mathbf{0}_{1 \times\left(L_0-1\right)}\right],
\end{equation}
where ${\bf R}_{\text {g}}(\tau )$ is the autocorrelation function of $g(t)$ sampled at $t=\tau T$. The vector $\bm\eta$ stands for the ideal zero-ISI sampled waveform after matched filtering.

Substituting (\ref{eq:convert}) into the objective function of (\ref{eq:problem}), we obtain
\begin{align}\label{eq:MSE0}
\text{MSE}(\bm\Theta,\!\bf G)&=\text{tr}\left({\bf T}({\bm\eta}){\bf R}_{\text s}{\bf T}^H({\bm\eta})\right)-2{\text{Re}}\left\{\text{tr}\left({\bf G}{\bf A}_{\epsilon}{\bf H}_{\text{eq}}{\bm\Theta}{\bf R}_{\text s}{\bf T}^H({\bm\eta})\right)\right\}\nonumber\\
&+\text{tr}\left({\bf G}{\bf A}_{\epsilon}{\bf H}_{\text{eq}}{\bm\Theta}{\bf R}_{\text s}{\bm\Theta}^H{\bf H}_{\text{eq}}^H{\bf A}_{\epsilon}^H{\bf G}^H\right)\!+\!\text{tr}\left({\bf G}{\bf R}_{\text v}{\bf G}^H\right),
\end{align}
where ${\bf R}_{\text s}\triangleq\mathbb{E}\{{\bf ss}^H\}$ and ${\bf R}_{\text v}\triangleq\mathbb{E}\{{\bf vv}^H\}$. By differentiating (\ref{eq:MSE0}) with respect to $\bf G$ and setting the derivative to zero, the optimal solution to $\bf G$ is obtained as
\begin{equation}\label{eq:equlizer}
{\bf{G}}={\bf T}(\bm{\eta }){\bf{R}}_{\text{s}}^H{{\bm{\Theta }}^H}{\bf{H}}_{{\rm{eq}}}^H{\bf{A}}_\epsilon ^H{\left( {{{\bf{A}}_\epsilon }{{\bf{H}}_{{\rm{eq}}}}{\bf{\Theta }}{{\bf{R}}_{\text{s}}}{{\bf{\Theta }}^H}{\bf{H}}_{{\rm{eq}}}^H{\bf{A}}_\epsilon ^H+{{\bf{R}}_v}} \right)^{ - 1}}.
\end{equation}
By substituting (\ref{eq:equlizer}) back to (\ref{eq:MSE0}), we have the expression of MSE in (\ref{eq:MSE}). \begin{align}\label{eq:MSE}
&\text{MSE}(\bm\Theta)\nonumber\\
&=\!-\text{2tr}\left({\bf T}({\bm\eta}){\bf R}_{\text s}^H{\bm\Theta}^H{\bf H}_{\text{eq}}^H{\bf A}_{\epsilon}^H\left({\bf A}_{\epsilon}{\bf H}_{\text{eq}}{\bm\Theta}{\bf R}_{\text s}{\bm\Theta}^H{\bf H}_{\text{eq}}^H{\bf A}_{\epsilon}^H\!\!+\!{\bf R}_{\text v}\right)^{-1}\!{\bf A}_{\epsilon}{\bf H}_{\text{eq}}{\bm\Theta}{\bf R}_{\text s}{\bf T}^H({\bm\eta})\!\right)\!+\text{tr}\left({\bf T}({\bm\eta}){\bf R}_{\text s}{\bf T}^H({\bm\eta})\right)\nonumber\\
&+\text{tr}\left({\bf T}({\bm\eta}){\bf R}_{\text s}^H{\bm\Theta}^H{\bf H}_{\text{eq}}^H{\bf A}_{\epsilon}^H\left({\bf A}_{\epsilon}{\bf H}_{\text{eq}}{\bm\Theta}{\bf R}_{\text s}{\bm\Theta}^H{\bf H}_{\text{eq}}^H{\bf A}_{\epsilon}^H+{\bf R}_{\text v}\right)^{-1}\!\!{\bf A}_{\epsilon}{\bf H}_{\text{eq}}{\bm\Theta}{\bf R}_{\text s}{\bm\Theta}^H{\bf H}_{\text{eq}}^H{\bf A}_{\epsilon}^H\left({\bf A}_{\epsilon}{\bf H}_{\text{eq}}{\bm\Theta}{\bf R}_{\text s}{\bm\Theta}^H\right.\right.\nonumber\\
&\left.\left.{\bf H}_{\text{eq}}^H{\bf A}_{\epsilon}^H+{\bf R}_{\text v}\right)^{-1}{\bf A}_{\epsilon}{\bf H}_{\text{eq}}{\bm\Theta}{\bf R}_{\text s}{\bf T}^H({\bm\eta})\right)+\text{tr}\left({\bf T}({\bm\eta}){\bf R}_{\text s}^H{\bm\Theta}^H{\bf H}_{\text{eq}}^H{\bf A}_{\epsilon}^H\left({\bf A}_{\epsilon}{\bf H}_{\text{eq}}{\bm\Theta}{\bf R}_{\text s}{\bm\Theta}^H{\bf H}_{\text{eq}}^H{\bf A}_{\epsilon}^H+{\bf R}_{\text v}\right)^{-1}{\bf R}_{\text v}\right.\nonumber\\
&\left.\left({\bf A}_{\epsilon}{\bf H}_{\text{eq}}{\bm\Theta}{\bf R}_{\text s}{\bm\Theta}^H{\bf H}_{\text{eq}}^H{\bf A}_{\epsilon}^H+{\bf R}_{\text v}\right)^{-1}{\bf A}_{\epsilon}{\bf H}_{\text{eq}}{\bm\Theta}{\bf R}_{\text s}{\bf T}^H({\bm\eta})\right)\!\nonumber\\
&=\underbrace {\text{tr}\left({\bf T}({\bm\eta}){\bf R}_{\text s}{\bf T}^H({\bm\eta})\right)}_{{\text{MSE}}_0}-\underbrace {{\text{tr}\left({\bf T}({\bm\eta}){\bf R}_{\text s}^H{\bm\Theta}^H{\bf H}_{\text{eq}}^H{\bf A}_{\epsilon}^H\left({\bf A}_{\epsilon}{\bf H}_{\text{eq}}{\bm\Theta}{\bf R}_{\text s}{\bm\Theta}^H{\bf H}_{\text{eq}}^H{\bf A}_{\epsilon}^H+{\bf R}_{\text v}\right)^{-1}{\bf A}_{\epsilon}{\bf H}_{\text{eq}}{\bm\Theta}{\bf R}_{\text s}{\bf T}^H({\bm\eta})\right)}}_{\overline{\text{MSE}}(\bm\Theta)}.
\end{align}
Then by safely dropping the constant term of MSE, i.e., MSE$_0$ in (\ref{eq:MSE}), the problem in (\ref{eq:problem}) is equivalent to the following maximization problem:
\setcounter{equation}{9}
\begin{align}\label{eq:problem1}
&{\mathop {\text{maximize}}\limits_{_{\bm\Theta}}}\quad\overline{\text{MSE}}(\bm\Theta)\nonumber\\
&\quad\quad\quad\text{s.t.}\quad(\text{5b}),(\text{5c}),
\end{align}
where $\overline{\text{MSE}}(\bm\Theta)={\text{tr}\left({\bf T}({\bm\eta}){\bf R}_{\text s}^{H/2}{\bf X}^H\left({\bf X}{\bf X}^H+{\bf R}_{\text v}\right)^{-1}{\bf X}{\bf R}_{\text s}^{1/2}{\bf T}^H({\bm\eta})\right)}$ with ${\bf X}\triangleq{\bf{A}}_\epsilon{\bf{H}}_{{\rm{eq}}}{\bm{\Theta }}{\bf R}_{\text s}^{1/2}$.
However, the equivalent problem in (\ref{eq:problem1}) is still non-trivial due to the following reasons. First, the objective function in (\ref{eq:problem1}) consists of fourth-order polynomials. Second, the optimization variable $\bm\Theta$ is required to be sparse and block-diagonal, i.e., ${\bm\Theta}={\bm\theta}\otimes{\bf I}_L$. Third, the non-zero elements in $\bm\Theta$ are constrained to be unit-modulus. To obtain a tractable solution, we develop an efficient algorithm based on the
MM framework in the following.
\subsection{Design of RIS Phase Shift Matrices}
The basic idea of the MM technique is to tackle a difficult problem by solving a series of simple subproblems iteratively. In particular, the objective of the subproblem approximates the original objective at the current point in each iteration. The key question for the application of the MM method is how to find a surrogate function such that the original maximization (minimization) problem can be well approximated by iteratively maximizing (minimizing) the surrogate function. For the optimization problem (\ref{eq:problem1}) at hand, the MM method can be employed as described in the rest of this subsection.

To start with, we first establish a lower bound (i.e., the surrogate function) of the objective function in (\ref{eq:problem1}) to facilitate the design of the algorithm. We exploit the fact that the function $f({\bf X}, {\bf Z}) = {\text {tr}}\left({\bf X}^H{\bf Z}^{-1}{\bf X}\right)$ is jointly convex in ${\bf Z\succ 0}$ and ${\bf X}$ [11]. Accordingly, the objective function of (\ref{eq:problem1}) is jointly convex in $\{{\bf X},{\bf P}\}$ by letting ${\bf P}\triangleq\left({\bf X}{\bf X}^H+{\bf R}_{\text v}\right)\succ {\bf 0}$.

Then, we commence with the following lemma that provides a suitable lower bound for the objective $\overline{\text{MSE}}(\bm\Theta)$ in (\ref{eq:problem1}).

\emph{Lemma 1:} For any feasible $\bm\Theta$ for (\ref{eq:problem1}) and given any feasible point ${\bm\Theta}_{(t)}$ at the $t$th iteration of the MM method, $\overline{\text{MSE}}(\bm\Theta)$ can be lower bounded by
\begin{equation}\label{eq:MSEq1}
\overline{\text{MSE}}(\bm\Theta)\geq\overline{\text{MSE}}({\bm\Theta}_{(t)})+2\text{Re}\left\{\text{tr}\left({\bf F}_{(t)}^H{\bf X}{\bf R}_{\text s}^{1/2}{\bf T}^H\right)\right\}-\text{tr}\left({\bf F}_{(t)}^H{\bf X}{\bf X}^H{\bf F}_{(t)}\right)+{\text {constant}},
\end{equation}
where ${\bf F}_{(t)}\triangleq\left({\bf X}_{(t)}{\bf X}_{(t)}^H+{\bf R}_{\text v}\right)^{-1}{\bf X}_{(t)}{\bf R}_{\text s}^{1/2}{\bf T}^H({\bm\eta})$ with ${\bf X}_{(t)}={\bf{A}}_\epsilon{\bf{H}}_{{\rm{eq}}}{\bm{\Theta }}_{(t)}{\bf R}_{\text s}^{1/2}$ and the equality is achieved when ${\bm\Theta}={\bm\Theta}_{(t)}$.
\begin{IEEEproof} $\overline{\text{MSE}}(\bm\Theta)$ is jointly convex in ${\bf X}$ and ${\bf P}$, therefore is lower bounded by its first order Taylor's expansion around $({\bf X}_{(t)}, {\bf P}_{(t)})$. Note that ${\bf X}_{(t)}={\bf{A}}_\epsilon{\bf{H}}_{{\rm{eq}}}{\bm{\Theta }}_{(t)}{\bf R}_{\text s}^{1/2}$ and ${\bf P}_{(t)}=\left({\bf X}_{(t)}{\bf X}_{(t)}^H+{\bf R}_{\text v}\right)$, the lemma is easily proved.\end{IEEEproof}
In order to maximize the lower bound in (\ref{eq:MSEq1}), we present the following lemma which allows us to acquire a closed-form solution of $\bm\Theta$ with a low computational complexity.

\emph{Lemma 2:} The function ${\text {tr}}\left({\bf Z}{\bf X}{\bf M}{\bf X}^H\right)$ is majorized by $-2 \text{Re}\left\{\text{tr}\left(\left(\lambda \bf{X}_{(t)}-\bf{Z} \bf{X}_{(t)} \bf{M}\right)^{H} \bf{X}\right)\right\}+\lambda\|{\bf X}\|_{F}^{2}$ +constant, given ${\bf M}\in\mathbb{C}^{N\times N}\succeq{\bf 0},{\bf Z}\in\mathbb{C}^{M\times M}\succeq{\bf 0}$, and any ${\bf X}_{(t)}\in\mathbb{C}^{M\times N}$ with $\lambda=\|{\bf M}\|_1\|{\bf Z}\|_1$.
\begin{IEEEproof} See the appendix.\end{IEEEproof}

By applying \emph{Lemma 2}, we obtain
\begin{equation}\label{eq:MSEq2}
-\text{tr}\left({\bf F}_{(t)}^H{\bf X}{\bf X}^H{\bf F}_{(t)}\right)\geq-\lambda_{(t)}\|{\bm\Theta}\|^2+\text{constant}+2\text{Re}\left\{\text{tr}\left(\left(\lambda_{(t)}{\bm\Theta}_{(t)}^H\!-\!{\bf R}_{\text s}^{H/2}{\bf X}_{(t)}^H{\bf F}_{(t)}{\bf F}_{(t)}^H{\bf A}_{\epsilon}{\bf H}_{\text{eq}}\right){\bm\Theta}\right)\right\},
\end{equation}
where $\lambda_{(t)}\triangleq\|{\bf R}_{\text s}\|_1\|{\bf H}_{\text{eq}}^H{\bf A}_{\epsilon}^H{\bf F}_{(t)}{\bf F}_{(t)}^H{\bf A}_{\epsilon}{\bf H}_{\text{eq}}\|_1$.

Combining (\ref{eq:MSEq1}) and (\ref{eq:MSEq2}), it yields a second minorization to $\overline {\text {MSE}}(\bm\Theta)$ as follows:
\begin{align}\label{eq:MSEq3}
\overline{\text{MSE}}(\bm\Theta) &\geq 2\text{Re}\left\{\text{tr}\left(\left(\lambda_{(t)}{\bm\Theta}_{(t)}^H+{\bf R}_{\text s}{\bf T}^H{\bf F}_{(t)}^H{\bf A}_{\epsilon}{\bf H}_{\text{eq}}-{\bf R}_{\text s}^{H/2}{\bf X}_{(t)}^H{\bf F}_{(t)}{\bf F}_{(t)}^H{\bf A}_{\epsilon}{\bf H}_{\text{eq}}\right){\bm\Theta}\right)\right\}+{\text {constant}}\nonumber\\
&\triangleq g_{\text{MSE}}\left({\bm\Theta}, {\bm\Theta}_{(t)}\right),
\end{align}
where we use the fact that $\|{\bm\Theta}\|^2=NK$. Note that to obtain a suboptimal solution to (\ref{eq:problem}), it suffices to iteratively solve the following problem:
\begin{align}\label{eq:problem2}
&{\mathop{\text{maximize}}\limits_{_{{\bm\Theta}}}}\quad g_{\text{MSE}}\left({\bm\Theta}, {\bm\Theta}_{(t)}\right) \nonumber\\
&\quad\quad\quad\text{s.t.}\quad(\text{5b}),(\text{5c}),
\end{align}
the optimal solution for which is given in the following lemma.

\emph{Lemma 3:} For any ${\bm\Theta}_{(t)}$, the problem (\ref{eq:problem2}) is solved by
\begin{equation}\label{eq:solution}
{\theta}_{i,j}=e^{-j{\text{arg}}\left({\text{tr}}\left({\bf B}_{(t)}[i,j]\right)\right)},
\end{equation}
where $\theta_{i,j}$ denotes the \!$j$th element of ${\bm\theta}_i$ and ${\bf B}_{(t)}[i,j]$ is a submatrix of ${\bf B}_{(t)}\triangleq\lambda_{(t)}{\bm\Theta}_{(t)}^H+{\bf R}_{\text s}{\bf T}^H{\bf F}_{(t)}^H{\bf A}_{\epsilon}{\bf H}_{\text{eq}}-{\bf R}_{\text s}^{H/2}{\bf X}_{(t)}^H{\bf F}_{(t)}{\bf F}_{(t)}^H{\bf A}_{\epsilon}{\bf H}_{\text{eq}}$ with columns from $\left(iN-N+j-1\right)L+1$ to $\left(iN-N+j\right)L$, for $i=1,\cdots,K$, and $j=1,\cdots,N$.
\begin{IEEEproof} By considering ${\bm{\Theta }} \triangleq\bm\theta\otimes {{\bf{I}}_L}$, $\bm\theta\triangleq{\left[ {\bm \theta _1^T, \cdots ,\bm \theta _K^T} \right]^T}$ and neglecting the constant terms that do not depend on $\bm\Theta$, problem (\ref{eq:problem2}) can be equivalently recast as
\begin{align}\label{eq:problem3}
&{\mathop{\text{maximize}}\limits_{_{{\theta}_{i,j}}}}~ \sum_{i,j}^{}2\text{Re}\left\{\text{tr}\left({\bf B}_{(t)}[i,j]\right){\theta}_{i,j}\right\}\nonumber\\
&\quad\quad\quad\text{s.t.}~\left| \theta _{i,j}\right| = 1,i\in\{1, \cdots K\},~j\in\{1, \cdots ,N\}.
\end{align}
Providing that $\{\theta_{i,j}\}$'s are independent variables, then (\ref{eq:problem3}) is maximized when the phases of $\{\theta_{i,j}\}$'s are aligned with those of \begin{small}$\left\{{\text{tr}}\left({\bf B}_{(t)}^H[i,j]\right)\right\}$\end{small}'s. Hence, (\ref{eq:solution}) is obtained.
\end{IEEEproof}
Now, we summarize the overall procedure to obtain a suboptimal solution to (\ref{eq:problem}) in Algorithm 1. We now give a detailed analysis of the convergence for Algorithm 1 by using the following proposition.

\emph{Proposition 1:} The objective value for the maximization problem in (\ref{eq:problem1}) is monotonically non-decreasing over iteration, i.e., $\overline{\text{MSE}}({\bm\Theta}_{(t)})\geq\overline{\text{MSE}}({\bm\Theta}_{(t-1)})$.
\begin{IEEEproof}It can be readily proved that
\begin{equation}\label{eq:pro1}
\overline{\text{MSE}}({\bm\Theta}_{(t)})\overset{(\text a)}\geq g_{\text{MSE}}\left({\bm\Theta}_{(t)}, {\bm\Theta}_{(t-1)}\right)\overset{(\text b)}\geq g_{\text{MSE}}\left({\bm\Theta}_{(t-1)}, {\bm\Theta}_{(t-1)}\right)\overset{(\text c)}\geq \overline{\text{MSE}}({\bm\Theta}_{(t-1)}),
\end{equation}
where (a) follows from (\ref{eq:MSEq3}), (b) holds because ${\bm\Theta}_{(t)}$ maximizes $g_{\text{MSE}}\left({\bm\Theta}, {\bm\Theta}_{(t-1)}\right)$, and (c) holds due to the equality $\overline{\text{MSE}}({\bm\Theta}_{(t)})= g_{\text{MSE}}\left({\bm\Theta}_{(t)}, {\bm\Theta}_{(t)}\right)$.\end{IEEEproof}

Owing to the above proposition and the optimality of $\bf G$ in (\ref{eq:equlizer}), the objective of problem (\ref{eq:problem}) is non-increasing after each iteration of Algorithm 1. Moreover, the objective value has a finite lower bound. Therefore, Algorithm 1 always converges. Furthermore, upon convergence, the first-order optimality properties of problem (\ref{eq:problem}) are also satisfied [12].

Concerning the computational complexity, it mainly involves the operations of matrix inversions and multiple matrix multiplications in each iteration. By considering Gaussian eliminations for
calculating the matrix inversions, the per-iteration computational complexity of Algorithm 1 amounts to the order of ${\cal O}\left((NKL)^3\right)$.
\begin{algorithm}[t]
\caption{Alternating Algorithm for Problem (\ref{eq:problem})}
\label{alg:Framwork}
\begin{algorithmic}[1] 
\STATE Set $t=0$, and initialize ${\theta}_{i,j}^{(t)},i=1,\cdots,K;j=1,\cdots,N.$
\REPEAT
\STATE${\bm{\Theta }}_{(t)} ={\left[{\theta}_{1,1}^{(t)},\cdots,{\theta}_{1,N}^{(t)},\cdots,{\theta}_{K,N}^{(t)}\right]^T} \otimes {{\bf{I}}_L}$;
\STATE${\bf X}_{(t)}={\bf{A}}_\epsilon{\bf{H}}_{{\rm{eq}}}{\bm{\Theta }}_{(t)}{\bf R}_{\text s}^{1/2}$;
\STATE${\bf F}_{(t)}=\left({\bf X}_{(t)}{\bf X}_{(t)}^H+{\bf R}_{\text v}\right)^{-1}{\bf X}_{(t)}{\bf R}_{\text s}^{1/2}{\bf T}^H({\bm\eta})$;
\STATE$\lambda_{(t)}=|{\bf R}_{\text s}\|_1\|{\bf H}_{\text{eq}}^H{\bf A}_{\epsilon}^H{\bf F}_{(t)}{\bf F}_{(t)}^H{\bf A}_{\epsilon}{\bf H}_{\text{eq}}\|_1$;
\STATE${\bf B}\left({\bm\Theta}_{(t)}\right)=\lambda_{(t)}{\bm\Theta}_{(t)}^H+{\bf R}_{\text s}{\bf T}^H{\bf F}_{(t)}^H{\bf A}_{\epsilon}{\bf H}_{\text{eq}}$
\STATE\quad\quad\quad\quad$-{\bf R}_{\text s}^{H/2}{\bf X}_{(t)}^H{\bf F}_{(t)}{\bf F}_{(t)}^H{\bf A}_{\epsilon}{\bf H}_{\text{eq}}$;
\STATE${\theta}_{i,j}^{(t+1)}=e^{-j{\text{arg}}\left({\text{tr}}\left\{{\bf B}[i,j]\right\}\right)},~i = 1, \cdots K;j = 1, \cdots ,N;$
\STATE$t \leftarrow  t+1$
\UNTIL{convergence}
\STATE${\bf G}={\bf T}({\bm\eta}){\bf R}_{\text s}^{H/2}{\bf X}_{(t)}^H{\left( {\bf X}_{(t)}{\bf X}_{(t)}^H+{\bf R}_{\text v} \right)}^{ - 1}.$
\end{algorithmic}
\end{algorithm}
\vspace{-5pt}
\section{Simulation Results}
In this section, we present numerical results to show the advantage of the proposed joint cooperative reflection and synchronization design compared with various schemes.

A uniform rectangular array is considered at the RIS with $N=N_xN_y$, where $N_x$ and $N_y$ denote the number of elements in the horizontal axis and vertical axis, respectively. For convenience, we adopt $N_x = 4$. Since the channel between the source and each RIS is line-of-sight (LOS) dominated according to measurement campaigns, we characterize it by a rank-one geometric channel model given as ${\bf f}_k=\sqrt N\lambda_k{\bm\alpha}_k(\vartheta _k^a,\vartheta_k^e)$,
where $\lambda_k\sim {\cal {CN}}(0,1)$ and $\vartheta _k^a (\vartheta_k^e)$ denote the complex gain and the azimuth (elevation) angle-of-arrival of the LOS path between ${\mathbb S}$ and ${\mathbb I}_k$, respectively. ${\bm\alpha}_k\in{\mathbb C}^{N\times 1}$ is the array response vector of ${\mathbb I}_k$, which is given in [13, (7)]. As for the channel between ${\mathbb I}_k$ and ${\mathbb D}$, we adopt the widely-used geometric model to depict the nature of high-frequency propagations for millimeter wave channel:
\begin{equation}\label{eq:mmwave}
\setlength\abovedisplayskip{1pt}
\setlength\belowdisplayskip{1pt}
\mathbf{h}_{k}^{\mathrm{H}}=\sqrt{\frac{N}{N_{p}^k}} \sum_{l=1}^{N_{p}^k} \lambda_{k}^{l}{\bm\alpha}_k^{H}\left(\phi_{k,a}^{l},\phi_{k,e}^{l}\right),
\end{equation}
where $N_p^k$ and $\lambda_k^l\sim {\cal {CN}}(0,1)$ represent the number of propagation paths and the complex gain of the $l$th path from ${\mathbb I}_k$ to ${\mathbb D}$, respectively. $\phi _{k,a}^l (\phi_{k,e}^l)$ denotes the azimuth (elevation) angle-of-departure of the $l$th path between ${\mathbb I}_k$ and ${\mathbb D}$.

In all simulations, $g(t)$ is chosen as a root-raised cosine pulse with an effective tail length $L_g=4$ and a roll-off factor of $0.3$ [14]. The symbol duration $T$ is set to $1/14$ ms. The length of the sequence is $L_0=12$ and the oversampling ratio $Q$ is set to $2$. The timing offsets are assumed to be uniformly distributed over the range $(-1,1)$. The signal-to-noise ratio (SNR) is defined as SNR$\triangleq\frac{E_s}{\sigma^2}$ with $E_s$ being the average transmitted signal power. For simplicity, the noise power is set to be $\sigma^2=1$ and the number of paths corresponding to different RISs are assumed to be the same, i.e., $N_p^1=N_p^2=\cdots=N_p^K=10$. The error tolerance factor is set to $10^{-4}$.

Fig. 2 illustrates the MSE of the proposed algorithm versus SNR. Solid lines correspond to the continuous phase cases while dashed lines represent the corresponding discrete cases. Here, we consider a baseline where the RISs are designed naively by assuming perfect timing synchronization, i.e., the blue line, by aligning the phase of the RISs with that of the cascaded channel as in [6], [7], i.e., ${\bm\theta}_k=e^{-j{\text{arg}}\left(\text{diag}({\bf h}_k^H){\bf f}_k\right)}$, $\forall k\in\{1,\cdots,K\}$. For the case with continuous phases, we find that the proposed design outperforms the baseline by orders-of-magnitude. This verifies the benefit of taking asynchronization into account in reflection optimization in the distributed RIS system, i.e., effectively mitigating the ISI and alleviating the signal-to-interference-plus-noise ratio penalty caused by asynchronization. The simulation results for discrete phases match well with the counterpart of continuous phases. Specifically, even with heavily quantized phase control, e.g., $B=2$, where $B$ denotes the number of quantization bits, the proposed algorithm incurs limited degradation.

In Fig. 3, we compare the MSE of different schemes versus the number of RISs, $K$, for both continuous and discrete phases, where SNR is set to 0 dB. It is observed that in both cases the performance gain achieved by the proposed design gradually increases as $K$ grows compared with the baseline case which over optimistically assumes perfect timing synchronization. This observation indicates that with more RISs deployed, the proposed joint RIS and equalizer design becomes more flexible to exploit the degrees-of-freedom brought by the multiple RISs for improving the system performance.
\begin{figure}
\begin{minipage}[t]{0.5\linewidth}
\centering
\includegraphics[width=3.5in]{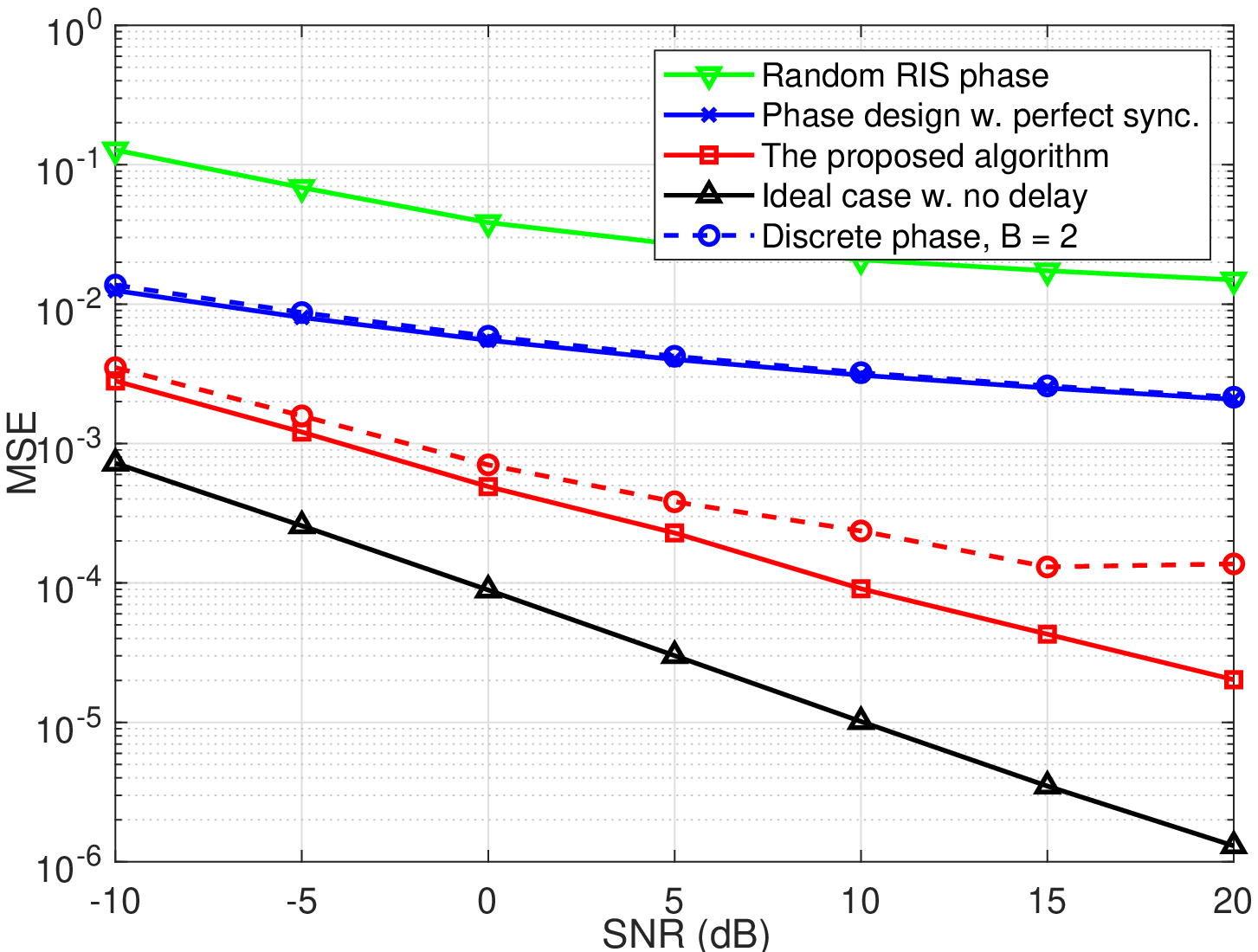}
\caption*{\small{Fig. 2 MSE versus SNR ($N=32,K=4$).}}
\label{fig:2}
\end{minipage}%
\begin{minipage}[t]{0.5\linewidth}
\centering
\includegraphics[width=3.5in]{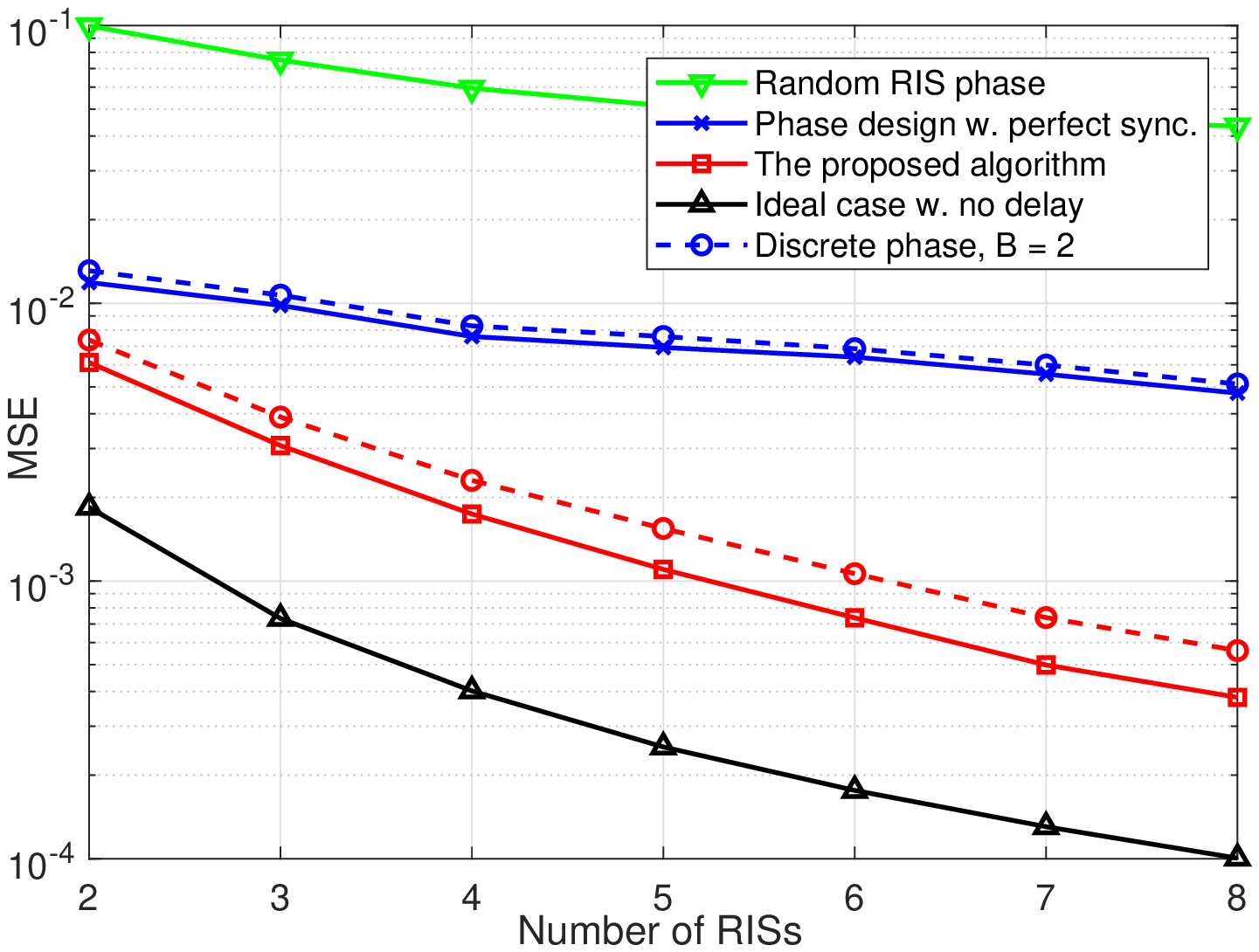}
\caption*{\small{Fig. 3 MSE versus the number of RISs ($N=16$).}}
\label{fig:3}
\end{minipage}
\end{figure}
\section{Conclusion}
We studied the MSE minimization problem for a distributed RIS system considering timing synchronization errors, where RIS phase shift matrices and the timing offset equalizer were jointly optimized. While the resultant optimization problem is nonconvex, we advocated a computationally-efficient algorithm to obtain a locally optimal solution based on the iterative MM framework. In each iteration, the original problem is solved with a  closed-form solution. Numerical results show that significant performance gains can be achieved by the proposed algorithm compared with various benchmark schemes.
\begin{appendix}
Since both ${\bf M}$ and ${\bf Z}$ are positive semidefinite matrices, we have $\|{\bf M}\|_1\geq\lambda_{\text{max}}({\bf M})$ and $\|{\bf Z}\|_1\geq\lambda_{\text{max}}({\bf Z})$. With $\lambda_{\text{max}}\left({\bf M}^T\otimes{\bf Z}\right)=\lambda_{\text{max}}({\bf M}^T)\lambda_{\text{max}}({\bf Z})=\lambda_{\text{max}}({\bf M})\lambda_{\text{max}}({\bf Z})$, it yields $\lambda\triangleq\|{\bf M}\|_1\|{\bf Z}\|_1\geq\lambda_{\text{max}}\left({\bf M}^T\otimes{\bf Z}\right)$. Consequently, the matrix $\lambda{\bf I}-{\bf M}^T\otimes{\bf Z}$ is constructed to be positive-definite. Then consider the following inequality:
\begin{equation}\label{eq:major}
\left\|\left(\lambda{\bf I}-{\bf M}^T\otimes{\bf Z}\right)^{1 / 2}\text{vec}({\bf X})-\left(\lambda{\bf I}-{\bf M}^T\otimes{\bf Z}\right)^{1 / 2}\text{vec}({\bf X}_{(t)})\right\|^{2}\geq 0,
\end{equation}
where the equality holds only when ${\bf X}={\bf X}_{(t)}$. Then it gives
\begin{align}\label{eq:lemma2}
\text{tr}\left(\mathbf{Z} \mathbf{X} \mathbf{M} \mathbf{X}^{H}\right)&=\text{vec}^{H}(\mathbf{X}) \text{vec}(\mathbf{Z} \mathbf{X} \mathbf{M})\overset{(\text a)}=\text{vec}^{H}(\mathbf{X})\left(\mathbf{M}^{T} \otimes \mathbf{Z}\right) \text{vec}(\mathbf{X})\nonumber\\
&\overset{(\text b)}\leq-2 \text{Re}\left\{\operatorname{vec}^{H}({\bf X}_{(t)})\left(\lambda \mathbf{I}-\mathbf{M}^{T} \otimes \mathbf{Z}\right) \operatorname{vec}\left(\mathbf{X}\right)\right\} \nonumber\\
&+\operatorname{vec}^{H}\!\left(\mathbf{X}_{(t)}\right)\left(\lambda \mathbf{I}-\mathbf{M}^{T} \otimes \mathbf{Z}\right)\text{vec}\left(\mathbf{X}_{(t)}\right)+\lambda \operatorname{vec}^{H}(\mathbf{X})\operatorname{vec}(\mathbf{X}) \nonumber\\
&\overset{(\text c)}=-2 \text{Re}\left\{\text{tr}\left(\left(\lambda \bf{X}_{(t)}-\bf{Z} \bf{X}_{(t)} \bf{M}\right)^{H} \bf{X}\right)\right\}+\lambda\|\mathbf{X}\|_{F}^{2}+\underbrace {\lambda\|{\bf X}_t\|_{F}^{2}-\text{tr}\left\{{\bf Z}{\bf X}_t{\bf M}{\bf X}_t^H\right\}}_{\text{constant}},
\end{align}
where (a) uses $\text{vec}\left({\bf{AXB}}\right)=\left( {{{\bf{B}}^T} \otimes {\bf{A}}} \right)\text{vec}({\bf X})$, (b) can be acquired by elaborating the inequality (\ref{eq:major}), and (c) is obtained by using similar manipulations as that in (a). By neglecting the constant term, we obtain the bound in \emph{Lemma 2}.
\end{appendix}

\end{document}